\documentstyle[11pt]{article}
\begin{document}
\title{Stability of inflationary solutions driven by a 
changing dissipative fluid}

\author{
Luis P Chimento\dag\,, 
Alejandro S Jakubi\dag\,
and Diego Pav\'on\ddag\
\\
\\
\dag\ {\small Departamento de F\'{\i}sica, Universidad de 
Buenos Aires, 1428~Buenos Aires, Argentina}\\
\ddag\ {\small Departament de F\'{\i}sica, Universidad Aut\'onoma 
de Barcelona, 08193 Bellaterra, Spain}\\
}

\maketitle
\date{\empty}
\begin{abstract}

In this paper the second Lyapunov method is used to study the stability of the
de Sitter phase of cosmic expansion when the source of the gravitational field
is a viscous fluid. Different inflationary scenarios related with reheating
and decay of mini-blackholes into radiation are investigated using an
effective fluid described by time--varying thermodynamical quantities.

\end{abstract}


\section{Introduction}
In recent years considerable attention has been paid to the 
bulk-viscosity driven inflationary scenario. This is only
natural since the effect of bulk viscosity in an expanding universe
is to reduce the equilibrium pressure. Therefore one may wish to know
whether this effect could be strong enough to render 
a large negative effective  pressure that leads to
inflation.

Fundamental strings can create an initial cosmological state either
of exponential or power law inflation followed 
by a smooth evolution towards the typical
Friedmann decelerated expansion dominated by the
stress-energy tensor of a radiation fluid. This connection 
is natural because the
string-driven inflationary expansion may arise due to 
the spontaneous
quantum production of fundamental strings on 
scales larger than the horizon, \cite{turok}
and a phenomenological bulk viscosity can be used to 
describe the effect of particle production \cite{JDB},
\cite{b3}. For some cosmological implications of
fundamental strings see also \cite{gv} and references
therein.

Inflationary scenarios have usually been associated 
with the dynamics of a spatially homogeneous scalar field 
such that its potential energy overpowered 
the kinetic energy, and the  equation of state the of vacuum 
$ p_{\phi} = -\rho_{\phi}$ was satisfied. If at the time of interest 
the scalar field dominated any other form of energy, then 
the cosmic scale factor increased exponentially with time \cite{kt}.
The fact that $\ddot{a}(t) > 0 $, where $a(t)$ is the scale factor, 
has several nice consequences, among others a scale invariant 
spectrum of initial density perturbations, the one more likely to
be compatible with observation \cite{pad}.

However inflation, either exponential or power-law, can in principle
be driven by  any mechanism that renders the total hydrostatic pressure
negative, such as bulk viscous pressure associated with non-adiabatic
expansion in FLRW universes -the effectiveness of this mechanism
has been discussed in the literature \cite{lpw}.
Usually in these models a mixture of 
relativistic and nonrelativistic (heavy) particles
is assumed at some early phase of cosmic expansion. There the bulk viscosity
can be very large and may drive inflation. After the decay of these heavy
particles bulk viscosity vanishes, terminating the inflationary phase, and
returning to the radiation dominated Friedmann universe \cite{PBJ}.

Very often a viscous pressure represents only a small perturbation to
the equilibrium (hydrostatic) pressure of the fluid. However, as is 
well known -see \cite{BLH} and \cite{ZLD}- the effect of particle
decay can be phenomenologically understood as a dissipative pressure,
and this one can be very large depending on how big the decay rate is.
This approach was put on a solid footing by Triginer {\it et al.} 
\cite{TZP}. 

Stability of the de Sitter solution in cosmological models where the
source of the metric is a dissipative fluid obeying some causal 
transport equation has been investigated by several authors. For 
the truncated version of the transport equation for  viscous 
pressure, stability has been
investigated in \cite{PBJ} and  \cite{Chi93}. In  \cite{mt},  
\cite{Chi96d}, \cite{Maa95} and \cite{cjm}  the full version of the 
transport equation -see (\ref{dpi}) below- was
adopted. The nonlinear full version \cite{rmvm} was used in 
\cite{Chi97}. The
rationale behind causal transport equations can be found in \cite{JCL},
for a short introduction see \cite{m2}.

Here we introduce a new approach by using the second method of
Lyapunov \cite{Ces} to examine the asymptotic
stability of the de Sitter solutions. 
This provides useful information on the dynamical behavior 
of the system, not only near the
stationary solutions but far away from them as well.

In section  2 the basic equations and Lyapunov's
criterion for stability are presented and applied to
different situations in which some or other of the three 
key parameters are held constant and the rest allowed to vary.
In section 3 the whole set of parameters are assumed
to vary. Finally section 4 summarizes the main findings 
of this work.  

\section{Asymptotic stability}
\label{sec2}
\subsection{General setting}
The spatially-flat
Friedmann-Lema\^{\i}tre-Robertson-Walker (FLRW) universe is described by the
metric

\begin{equation} \label{ds}
ds^2=-dt^2+a^2(t)\left(dx^2+dy^2+dz^2\right)
\end{equation}

Because of the spatial isotropy
and homogeneity assumptions, velocity gradients causing shear 
viscosity and temperature gradients leading to heat transport are 
absent. The only possible dissipative term corresponds to
the bulk viscosity (which as mentioned before may
be interpreted as the effect of particle production). So the
energy-stress tensor of this viscous fluid is given by

\begin{equation}
T^{ab} = \rho u^{a}u^{b}
+ \left(p+\pi\right) h^{ab}
\label{Tik}
\end{equation}

\noindent
where $\rho$ is the total energy density
of the cosmic fluid, $p$ its equilibrium pressure, $\pi$ the
dissipative scalar pressure,
$u^{a}$ the four-velocity, normalized so that
$u^{a} u_{a} = - 1$ and the tensor
$h^{ab}=g^{ab} + u^{a}u^{b}$ projects any tensorial quantity
into the hypersurface orthogonal to $u^{a}$.

The corresponding Einstein field equations read
\begin{equation} 
3H^{2}  = \kappa \rho
\label{ZPM28}  
\end{equation}

\begin{equation}
\dot{H}  = - {\kappa\over 2}\left(\rho + p + \pi\right)
\label{ZPM29}
\end{equation}

\noindent where  $H \equiv \dot{a}/a$ is the Hubble function, 
and $\kappa$ is Einstein's gravitational constant. 
We use units such that $c = k_{B} =
\hbar=1$, so that $\kappa=8\pi/M_{P}^2$, where $M_{P}$ is 
the Planck mass. An over-dot denotes derivative with respect to cosmic time.

As is well known, the dissipative pressure  obeys the causal 
evolution equation \cite{d1}, \cite{m2}
\begin{equation}
\pi \left[ 1 + \zeta \left( \frac{\tau}{2 T \zeta} u^{a} \right)_{;a} \right]
+ \tau \dot{\pi} = - 3 \zeta H 
\label{dpi}
\end{equation}  
where $\zeta$ indicates the phenomenological coefficient of bulk 
viscosity, $T$  the temperature of the cosmic fluid,
and $\tau$ the relaxation time associated to the dissipative pressure.
Usually the latter is given by the kinetic theory
of gases or by a fluctuation-dissipation theorem or both \cite{jpa}.
Provided the divergence on the left hand side is small 
the last equation can be approximated by
\begin{equation}
\pi + \tau\dot{\pi}  =  - 3\zeta H 
\label{dpi2}
\end{equation}

\noindent
In view of equation 
(13) -see below-  and the frequently made assumption that 
$\zeta \propto \rho$, which is very natural for radiative-like
fluids \cite{weinberg}, \cite{jpa},
the above approximation may be justified
for expanding regimes with  $H^{2} \gg \dot H$, something to be
expected at the commencement of the Universe expansion. In any case
the two above equations meet the requirements of causality and 
stability to be fulfilled by any physically acceptable transport
equation \cite{hl}.
At this point it is expedient to introduce the polytropic index

\begin{equation} \label{gamma}
\gamma \equiv 1 + {p\over\rho}
\end{equation}
which in general depends on time. This is obvious for a 
cosmic fluid consisting of  a mixture of massless and massive 
particles, as these two components redshift at different rates 
and consequently $\gamma$ varies with time. 

This set of equations combines to 
\begin{equation} \label{dH1}
\ddot H+3\gamma H\dot H+\tau^{-1}\left[\dot H+\frac{3}{2}
\left(\gamma+\tau\dot\gamma\right)H^2-\frac{3}{2}\zeta H\right]=0
\end{equation}

\noindent
and the latter can be recast as

\begin{equation} \label{d2}
\frac{d}{dt}\left[\frac{1}{2}{\dot H}^2+V(H)\right]=D(H,\dot H)
\end{equation}

\noindent
Here, the left hand side is the time derivative of a Lyapunov function
(see the appendix) with

\begin{equation} \label{V}
V(H)=\frac{\zeta}{2\tau}
\left(\frac{\gamma+\tau\dot\gamma}{\zeta}H^3-\frac{3}{2}H^2\right)
\end{equation}

\noindent
and

$$
D(H,\dot H)=-\left(3\gamma H+\tau^{-1}\right)\dot H^2+
\frac{1}{2}\tau^{-2}\left(\tau\dot\gamma-\dot\tau\gamma+
\tau^2\ddot\gamma\right)H^3
$$

\begin{equation} \label{D}
+\frac{3}{4}\tau^{-2}\left(\zeta\dot\tau-\dot\zeta\tau\right)H^2
\end{equation}

\noindent
Assuming that the ratio $\zeta/\tau$ is bounded
$V(H)$ has two extrema (see Fig. 1), a maximum at
$H=0$, and a minimum at

\begin{equation} \label{H1}
H=\frac{\zeta}{\gamma+\tau\dot\gamma}
\end{equation}

\noindent
Inserting $\zeta$ from (\ref{H1}) in (\ref{D}) we get

$$
D = -\left(3\gamma H+\tau^{-1}\right)\dot H^2-
\frac{3}{4}\tau^{-1}\left(\gamma+\tau\dot\gamma\right)H^2\dot H
$$
\begin{equation} \label{D1}
-\frac{1}{4}\tau^{-2}\left(\tau\dot\gamma-\dot\tau\gamma+\tau^2\ddot\gamma
\right)H^3
\end{equation}

\noindent
In the simplest case when the functions $\gamma$, $\tau$ and $\zeta$ are
constants, $D$ is seminegative definite in a neighborhood of the de Sitter
solution $H_1=\zeta/\left(\gamma+\tau\dot\gamma\right)$ and the Lyapunov function has a time independent upper
bound for large times, implying the stability of this solution \cite{Ces}. On
the other hand, when $H$ is slowly varying we can neglect $\dot H$ in
(\ref{D1}); thereby $D$ reduces to

\begin{equation} \label{D2}
D\approx -\frac{1}{4 \tau^{2}} \left(\tau\dot\gamma-\dot\tau\gamma+\tau^{2}
\ddot\gamma\right)H^3
\end{equation}

\noindent Provided $D$ is negative definite, and taking into account that the
Lyapunov function has an infinitesimal upper bound in the neighborhood of
point $(H_1,0)$ of the phase space $(H,\dot H)$, the
de Sitter solution (\ref{H1}) is asymptotically stable \cite{Ces}.

In contrast with the usual inflationary scenarios driven by a
large effective cosmological constant necessarily accompanied 
by strong supercooling and reheating, dissipative processes in 
our scheme reheat the medium gently.
In the following, several simple cases when either
$\gamma$ or $\tau$ remain constant are investigated. 

\subsection{Variable $\tau$}

In the first instance 
(\ref{H1}) is asymptotically stable when $\tau$ is
a decreasing function, while in the second case stability occurs when
$\gamma$ increases slowly. Thus, for instance, if $\gamma$ is a constant and
$\tau$ is a function that decreases in a first stage and then increases, we
have that there is a first period of exponential inflation with
$H=\zeta/\gamma$, and then a graceful exit.

For instance, using the relationship for the speed $v$ of the
dissipative signal

\begin{equation} \label{zetatau}
\frac{\zeta}{\tau}=v^2\gamma\rho
\end{equation}

\noindent
derived in \cite{m2}, we obtain  

\begin{equation} \label{tau}
\tau=\frac{1}{3v^2H}
\end{equation}

\noindent 
and therefore  $\tau$ will present a minimum provided $v^{2}$ has a
maximum. In this case it may be said that the dissipative effect 
both drives an inflationary stage and then causes the exit from it.

Another way to implement this scenario is as follows. Assume the cosmic
fluid is modelled by a mixture of radiation and heavy particles 
(or massive modes of fundamental strings) that decay at a very high
rate into (more stable) lighter particles (less massive modes),
with high or moderate multiplicity. Since the relaxation
time for the interaction between radiation and massive particles is
on very general grounds
given by $\tau \approx (n \sigma)^{-1} $, where $n$ denotes the 
number density of massive particles  and $\sigma$ the interaction
cross-section, which can be constant, there will be two stages.
The first one corresponds to the decay of the more massive particles;
there (despite the expansion) $n$ will augment and $\tau$ decrease 
accordingly. The second stage will commence when most of these 
particles have decayed; then the expansion will make $n$ decrease 
and correspondingly $\tau$ will increase \cite{ZPM}.

\subsection{Variable $\gamma$ and decay of massive particles}

This second case will be useful to model the cosmic evolution
during the inflationary
period by allowing massive (dust) particles to decay into 
relativistic particles \cite{D2}. Thus one may have 
simultaneously inflation and reheating.

Exit from inflation occurs when particle production 
ceases and the fast dilution of relativistic particles 
makes $\gamma$  decrease again.
An interesting model for the reheating-inflationary 
stage arises when $\tau$
and $\zeta$ are constant but  $\, \gamma $ varies in such a way 
that $H_1$ remains constant. 
Using (\ref{D}) we
see that $D$ is semidefinite negative and therefore 
the de Sitter stage is stable. In this case $\gamma(t)$ 
takes the form
\begin{equation} \label{gammat}
\gamma(t)=\gamma_0+C_2 e^{-t/\tau}
\end{equation}

\noindent
and $H_1=\zeta/\gamma_0$.  It is straightforward to implement
a cosmological model in which $\gamma=1$ at $t=t_{1}$ 
and $\gamma=4/3$ at $t=t_{2}$, with $t_{2}>t_{1}$

\begin{equation} \label{gammat2}
\gamma(t)=1+\frac{e^{-t/\tau}-e^{-t_1/\tau}}
{3\left(e^{-t_2/\tau}-e^{-t_1/\tau} \right)} 
\end{equation}

\noindent
The evolution of $\gamma(t)$ depends on the relationship between 
the timescales $\tau$ and $t_{2}-t_{1}$. When $\tau\ll t_{2} - t_{1}$ 
the relativistic stage is reached very quickly.
This corresponds to a universe initially dominated by 
very massive (dust) particles, that spontaneously decay 
into radiation at a high initial rate \cite{D2}.
On the other hand, when $\tau\gg t_2-t_1$, the growth of $\gamma(t)$
becomes milder.

Since the detailed behavior of a dissipative relativistic 
fluid in a FLRW universe strongly depends on the thermodynamical
properties of the cosmic fluid and generally these are poorly known,
we are led to use the Boltzmann gas as the cosmological medium to
obtain analytical results. Thus, while a Boltzmann gas is not a 
realistic model for the cosmological fluid in the actual Universe, 
it is a fluid for which the thermodynamical properties are 
well enough established by relativistic kinetic theory to allow us 
to build up precise models. The thermodynamic properties of the 
Boltzmann gas may be described by the dimensionless inverse 
temperature $ z =m/T$, the relativistic chemical
potential $\alpha$ and a constant 
$A_0=m^{4}g_{*}/(2\pi^{2})$, 
where $m$ is the particle mass and $g_{*}$ the
spin weight of the fluid particles. The ideal gas law 
has the form

\begin{equation} \label{p}
p=A_0e^{-\alpha}\frac{K_2(z)}{z^2}
\end{equation}

\begin{equation} \label{rho}
\rho=A_0\left[\frac{K_1(z)}{z}+3\frac{K_2(z)}{z^2}\right]
\end{equation}
where $K_{n}$ are modified Bessel functions of 
the second kind.
\noindent
Then, inserting (\ref{p}) and (\ref{rho}) in (\ref{gamma}), and assuming a
vanishing chemical potential, we obtain

\begin{equation} \label{gammabeta}
\gamma(z)=1+\frac{K_2(z)}{z K_1(z)+3K_2(z)}
\end{equation}

Equating (\ref{gammat2}) and (\ref{gammabeta}) 
we can describe the continuous 
process of decay of mini-black holes from  
$t = t_{1}$ when the black hole energy density dominates
the Universe, until $t_{2}$ when the black holes have
completely evaporated away 
and the Universe is radiation-dominated. In so doing 
we are implicitly assuming
that all the black holes have the same mass and therefore the same
temperature, and that this one  equals the temperature of the 
massless component of the cosmic fluid at the beginning of the
evaporation. Scenarios in which this situation may occur have 
been reported in the literature -see for instance \cite{Kodama}, 
\cite{BCK} and \cite{Linde}. In \cite{Kodama} black holes
are formed as consequence of the first order phase transition 
from the false to the true quantum vacuum; in \cite{BCK}
because of collisions between bubbles of the new phase, and in
\cite{Linde} by quantum fluctuations at the end of hybrid 
inflation.
In all three cases these abundantly produced mini-black holes
dominate the Universe and their subsequent explosive 
evaporation into lighter particles can be modeled as a 
dissipative pressure.

Following \cite{Lima} the equation for the evolution of the 
temperature of the radiation fluid is

\begin{equation} \label{dT}
\frac{\dot{T}}{T} =  \frac{9 H^2 \zeta }{T \partial\rho/\partial T}
-3H\frac{\partial p/\partial T}{\partial{\rho }/\partial T}
\end{equation}

\noindent
Inserting (\ref{p}), (\ref{rho}) and (\ref{gamma}) in (\ref{dT}) we obtain

\begin{equation} \label{dz}
\frac{z'}{z}=\frac{12 K_2(z)+3z K_1(z)-B z^2}{12 K_2(z)+5z K_1(z)+z^2 K_0(z)}
\end{equation}

\noindent where $B=9H\zeta/A_0$ and $'\equiv d/Hdt$ . From this equation it is
easy to see that $z'$ is negative for large $z$, which is in accord with the
reheating scenario of above (see Fig. 2).

We can explicitly obtain the time dependence of this temperature near 
$t_{1}$ and $t_{2}$. 
Expanding (\ref{gammat2}) about $t = t_{1}$ and 
(\ref{gammabeta}) for $z \to \infty$, we obtain 
$T \propto t-t_{1}$. In the opposite limit 
(i.e. $t\to t_{2}$ and $z \to 0$) it follows
that $T \propto (t_{2}-t)^{-1/2}$.
Thus -as found in similar scenarios \cite{ZD1}-
the production of relativistic particles at the final stage of 
the black holes evaporation is accompanied by  a huge increase 
of the temperature of the cosmic fluid. As for the temperature
of the black hole component we must say that this is essentially
zero since the black holes behave as a dust fluid (only that its
``particles" emit radiation and (simultaneously) absorb 
the ambience fluid). Nevertheless, one may choose
to ascribe a temperature to each individual mini-black hole of mass 
$ M $ by the Hawking relationship 
$T_{bh} \propto M^{-1}$ -see \cite{hw}. The evolution of this 
temperature is governed by the sum of two terms. One of them comes
from  the black hole evaporation $\propto M^{-4} $ \cite{hw}, and 
the other one comes from the accretion. The latter term 
is more complicated and not of much interest for our purposes here. 
In any case, the fate of the black holes (assuming they do not leave 
any stable relic behind) is their complete disappearance by yielding
their whole mass to the radiation fluid.

The entropy production per unit volume in the latter
is given by \cite{m2}

\begin{equation} \label{dS}
\dot S= \frac{\pi^2}{\zeta T}=\frac{9\gamma^2\zeta^3}{\gamma_0^4 T}
\end{equation}

\noindent
It has the limiting behavior

\begin{equation} \label{dS1}
\dot S \simeq \frac{3^7\tau\left(1-e^{-x}\right)^5}{m\left(3e^{-x}-2\right)^4
\left(t-t_1\right)}, \qquad t\to t_1
\end{equation}

\begin{equation} \label{dS2}
\dot S\simeq\frac{6^{9/2}\zeta^3\left(e^{-x}-1\right)^4\sqrt{t_2-t}}
{m\sqrt{\tau}\left(3e^{-x}-2\right)^4\sqrt{e^x-1}},
\qquad t\to t_2
\end{equation}

\noindent
where $x=(t_2-t_1)/\tau$. Thus we see that the entropy production rate 
is very high at the beginning of the evaporation, while it
decreases sharply at the final stage of this process. At first sight
this may seem counter-intuitive if one has in mind that the final stage
of black hole evaporation is explosive (when one adheres to the Hawking
picture as we do). However, our assumption of $\zeta $ = constant
implies that the aforementioned accretion renders the evaporation rate
much milder. The net rate of radiation particle production per
mini-black hole and unit of volume roughly varies as 
$(\rho + p)^{-1}$, where in this case $\rho $ and $p $ refer
to the radiation fluid only \cite{ZD2}. 

We note that the opposite process, i.e. the one in which
the accretion of radiation by the black holes overpowers the
evaporation of the latter, and as a consequence the radiation is
entirely eaten up by the black holes, is ruled out by the second
law of thermodynamics \cite{ZD1}.

\subsection{Variable $\gamma$ and $\tau$}

Assuming now that both $\gamma$ and $\tau$ change in time, 
with a time scale for $\gamma$ much larger than $\tau$, 
we may neglect the term $\tau^{2}\ddot\gamma$ in equation 
(\ref{D2}), whence we are left with
$D=-\left(1/4\right)\left(\gamma/\tau\right)^{.}H^{3}$. 
Thus, the exponential inflation is stable provided 
$\gamma/\tau$ is an increasing function, and
whenever it begins to decrease, the de Sitter stage ends.

The solution $H=0$ corresponds to the Minkowski solution.
To analyse its stability we 
linearize equation (\ref{dH1}) about it, 
\begin{equation} \label{delta}
\ddot H+\tau^{-1}\dot H-\frac{3}{2}\zeta\tau^{-1}H=0
\end{equation}

\noindent
The roots of the characteristic polynomial are

\begin{equation} \label{lambda}
\lambda_{\pm}=\frac{1}{2\tau}\left(- 1\pm\sqrt{
1 + 6 \zeta\tau}\right)
\end{equation}

\noindent 
Since $\zeta$ and $\tau$ are positive definite quantities it 
follows $\lambda_{-}<0<\lambda_{+}$ and therefore the
Minkowski solution is unstable. There exists,
however, a one-parameter family of solutions
that approaches a flat spacetime solution at large times. 
There is in addition a one-parameter family of solutions that 
starts from a Minkowski spacetime in the far past 
and evolves towards a stable de Sitter solution. The 
time-reversal of the latter is nonsingular and 
corresponds to a spatially flat universe.

\section{General case}

In this section we consider the de Sitter solution (\ref{H1}) and
assume that $\gamma, \tau$ and $\zeta$ are arbitrary functions. 
This more general 
situation may occur during the decay of massive particles into 
lighter ones and also during the decay of four-dimensional fundamental
strings into massive and massless particles -admittedly this second
possibility is more speculative. The differential equation to 
solve is  

\begin{equation} \label{dgamma}
\gamma+\tau\dot\gamma=\frac{\zeta}{H_1}
\end{equation}

We introduce a new dimensionless independent variable
$\mbox{d}\eta=\mbox{d}t/\tau$. The reason for using a dimensionless equation is that the
equilibrium point of the differential equation (\ref{dgamma}), describing
exponential inflationary models, will represent self-similar cosmological
models. Then the general solution reads

\begin{equation} \label{gammaeta}
\gamma=\frac{1}{H_1}e^{-\eta}\left(\int \mbox{d}\eta \, 
\zeta e^{\eta}+C\right)
\end{equation}

\noindent where $C$ is an arbitrary integration constant. We will use this
result to present two simple models. First we consider that 
$\zeta=\zeta_0$, a constant. Thereby

\begin{equation} \label{gammaeta2}
\gamma(\eta)=\frac{1}{H_1}\left(\zeta_0+C e^{-\eta}\right)
\end{equation}

\noindent
A natural choice is $\eta=\nu\ln t$, $\nu$ being a positive constant, 
that represents a
linear dependence between $\tau$ and the cosmological time $t$. 
Then we obtain

\begin{equation} \label{gammat3}
\gamma(t)=\gamma_0\left(1+\frac{\bar C}{t^{\nu}}\right)
\end{equation}

\noindent
with $\gamma_0=\zeta_0/H_1$.
This expression is monotonic increasing (decreasing) for $\bar C<0$
($\bar C>0$). For
the case that $\bar C=(t_1 t_2)^\nu/(4t_2^\nu-3t_1^\nu)$ and
$\gamma_0=(4t_2^\nu-3t_1^\nu)/(3(t_2^\nu-t_1^\nu))$, this model describes
decay of massive particles into radiation beginning at $t_1$ and finishing at
$t_2$.

Now inserting (\ref{gammat3}) into (\ref{zetatau}) we obtain an expression for
the dissipative contribution to the speed of sound

\begin{equation} \label{v}
v^2=\frac{\nu }{3H_1}\frac{t^{\nu-1}}{t^\nu+\bar C}
\end{equation}

\noindent This is a monotonic decreasing function that in the limit
$t\to\infty$ behaves as $v^2\sim \nu/(3H_1t)$. Thus we require that
$v(t_1)\le 1$.

Considering now that $\zeta=\zeta_0 e^{-\eta}$, the expression
(\ref{gammaeta}) gives

\begin{equation} \label{gammaeta3}
\gamma(\eta)=\frac{\zeta}{H_1}\left(\eta+C\right)
\end{equation}

\noindent
In order to simplify the calculations  we restrict to constant
$v$; in this case we obtain

\begin{equation} \label{teta}
\Delta t=\frac{1}{3H_1v^2}\ln\left|\eta+C\right|
\end{equation}

\noindent
In order to have $ \Delta t > 0$, we choose $\eta+C>0$;
then

\begin{equation} \label{gammat4}
\gamma(\Delta t)=\gamma_0\exp\left(C+3H_1v^2\Delta t-e^{3H_1v^2\Delta t}
\right)
\end{equation}

\noindent This expression has a maximum at $\Delta t=0$, and assuming that
$\gamma ( 0 )=4/3$ we find that there is a phase of decay into radiation
starting when $\gamma=1$ at $3H_1v^2\Delta t=-0.8678$ that reaches a
relativistic gas state at $\Delta t=0$. After that there is a condensation
phase back into nonrelativistic matter that ends at $3H_1v^2\Delta t=0.6736$
when $\gamma$ returns to $1$. A scenario compatible with the 
latter phase is the quantum tunneling of radiation into 
black holes \cite{GPY}. This may arise very naturally 
because of the instability of the hot radiation against 
spontaneous condensation \cite{kapusta}. (It is altogether 
different from the whole disappearance of the radiation 
by black hole accretion).

During this period both the viscosity coefficient

\begin{equation} \label{zetat}
\zeta(t)=\zeta_0\exp\left(C-e^{3H_1v^2\Delta t}\right)
\end{equation}

\noindent
and the relaxation time

\begin{equation} \label{taut}
\tau(t)=\frac{1}{3H_1v^2}e^{-3H_1v^2\Delta t}
\end{equation}

\noindent
are monotonic decreasing functions.

Again we may interpret this behavior in terms of a 
two-fluid model, where the
viscosity coefficient arises because of the particle production 
process from the decay of massive nonrelativistic particles into
light ones. 
Shortly after the beginning of the decay the particle production 
rate is large and the energy density of the fluid
becomes dominated by the light component. Later on, as the
decay rate slows down the effect of adiabatic dilution by the fast
exponential expansion of the universe turns out to be  more important. 
Therefore the nonrelativistic particles dominate again, since their number 
density goes down as $a^{-3}$, while the relativistic component goes 
down at the faster rate of $a^{-4}$.

\section{Conclusions}
With the help of the Lyapunov method we  have studied the 
stability of cosmic inflationary expansions 
driven by a dissipative fluid whose transport equation is
of causal type. We required
a slow transition from the symmetric to the broken phase. This 
transition may be even quasistatic. Exponential inflation 
occurs during the transition whenever the ratio
$\zeta/\left(\gamma+\tau\dot\gamma\right)$ 
remains constant, or at most varies slowly in time. 
The behavior of the scale factor changes gently 
from exponential expansion to Friedmannian ($ a(t) \propto t^{n}$,
with $0< n <1$) once the heavy particles have either decayed 
or become sufficiently diluted, in this way rendereing the 
viscosity negligible.

We have found that the de Sitter solution
is asymptotically stable for a wide set of reasonable fluid 
quantities of the dissipative cosmological medium. Thus the
condition for inflation appears natural, and
the extremely fast supercooling followed by an
intense reheating, so frequent in the literature, is avoided.

To obtain analytic results for the continuous decay processes of decay of
mini-black holes we have modeled the fluid by a Boltzmann gas. Thus we are
able to obtain the time evolution of the temperature from non-relativistic to
the ultrarelativistic regime reflecting in an increase of the adiabatic index
from $1$ to $4/3$. In this way a reheating phase is shown to occur
simultaneously with exponential inflation.

Concerning the structural stability of the de Sitter solution relative to the
spatial geometry, we have  neglected the effect of curvature in comparison with
the energy density and effective pressure during an exponential inflationary
stage. In the case of non-causal bulk viscosity, where $\pi$ is
algebraically determined by $H$, it is relatively straightforward to
investigate this question \cite{b1,JDB,b3}. By contrast, in the causal theory
$\pi$ is no longer algebraically determined by $H$ but satisfies a transport
equation that couples it differentially to the expansion. Curvature introduces
the scale factor explicitly into the Friedmann equation, and makes it much
harder to decouple the equations. Thus the effect of curvature in the
causal case is far more difficult to determine in general. This will be
the subject of future work.

\section*{Acknowledgments}
We are grateful to Roy Maartens for critically reading the manuscript
and useful comments. This research has been partially supported by 
the Spanish Ministry of Education under grant PB94-0718.
LPC and ASJ thank the University of Buenos Aires for partial 
support under project EX-260.

\section*{Appendix}

The stability of a solution of equation (\ref{dH1}) can also  be studied
by linearization about this solution. However, 
this procedure leads to a linear differential equation for the 
perturbation with time-dependent coefficients,
through three unspecified functions: $\zeta(t)$, $\gamma(t)$ and $\tau(t)$. 
Consequently it is extremely difficult 
to determine the behavior of the perturbation for large
times, as it requires to calculate the perturbation as a functional of these
coefficients. This is why we have chosen the second method of Lyapunov.
This method has led us to some general qualitative results about the 
stability of the de Sitter solution.

For a time-dependent Lyapunov function we have used the following theorems
\cite{Lya}:

\begin{enumerate}

\item
{\em If a function $V$ exists which is defined and whose derivatives $V'$ is a
semidefinite function whose sign is contrary of that of $V$, then the solution
$x=0$ of}

$$
x_i'=f_i\left(t,x_1,\cdots,x_n\right),\qquad
i=1,2,\cdots,n,
\eqno (A1)
$$

\noindent
{\em is stable}.

\item
{\em If a function $V$ exits which is definite, and has an infinitesimal upper
bound, if the derivative $V'$ is also a definite function whose sign is
contrary to that of $V$, then the solution $x=0$ of} (A1)
{\em is asymptotically stable.}
\end{enumerate}

\noindent
In our case $x \equiv \left(H-H_1,\dot H\right)$ and the Lyapunov 
function is

$$
\frac{1}{2}\dot H^2+
\frac{\zeta}{2\tau}\left(\frac{\gamma+\tau\dot\gamma}{\zeta}H^3-
\frac{3}{2}H^2\right)
$$

\noindent
which clearly satisfies the hypotesis of the theorem provided that 
$\zeta/\tau$ is bounded (see equation (15)), in a neighbourhood of 
the de Sitter solution $H_1=\zeta/\left(\gamma+\tau\dot\gamma\right)$.

\newpage

\noindent
{\Large \bf Figure Captions}

\bigskip \noindent Figure 1.
Potential $V(H)$ of the
equivalent mechanical system defined by equation (\ref{V}).

\bigskip \noindent Figure 2.
Plot of the dimensionless time derivative of the inverse temperature $z'$
for $B=0.1$.
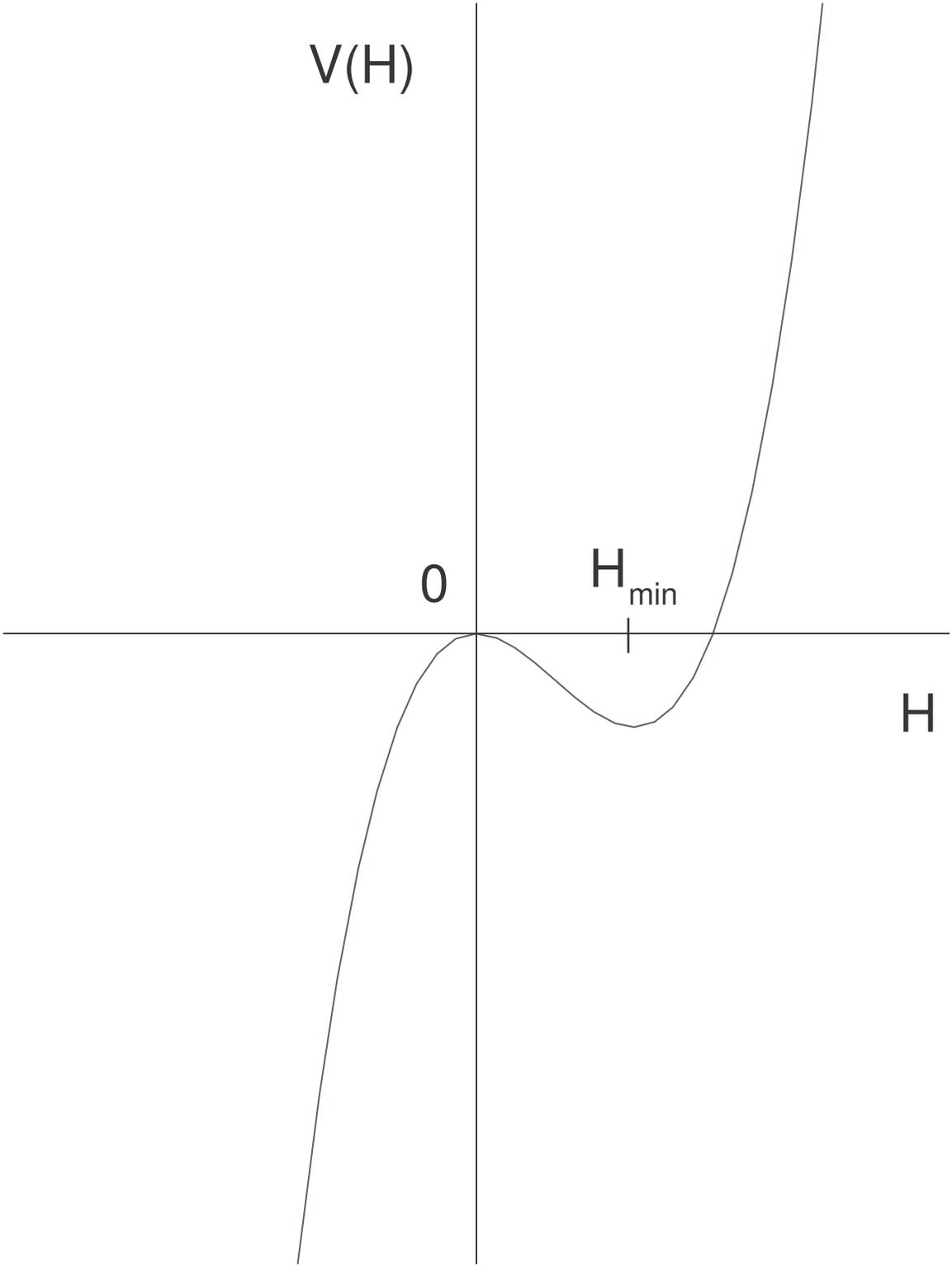
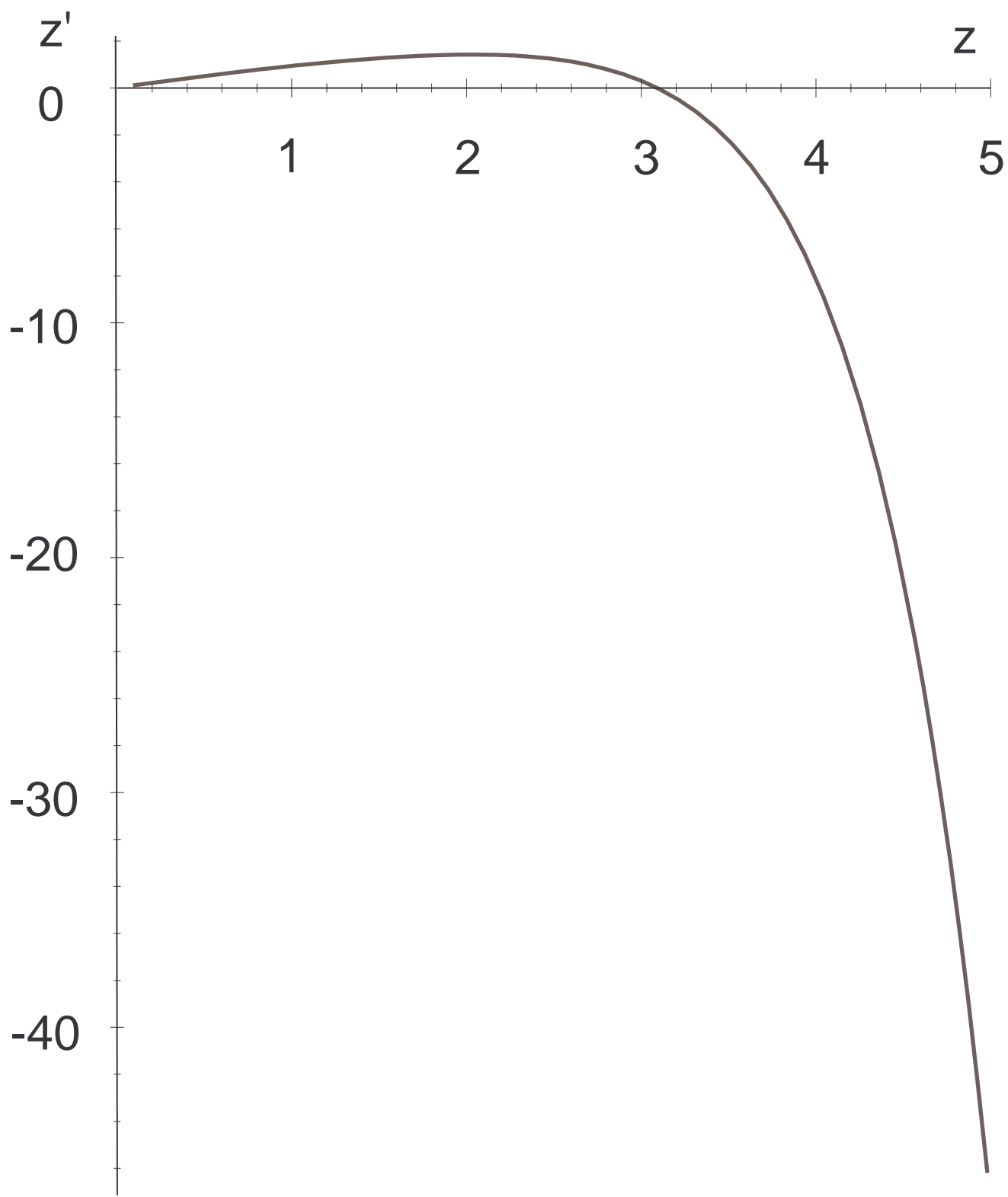
\end{document}